\documentclass[12pt]{article}
\usepackage{amssymb,amsmath}
\usepackage{amsfonts}
\usepackage[dvips]{graphicx}
\usepackage{latexsym,graphicx,tikz,setspace,wasysym,float,subfig}

\newcommand{\nn}{\nonumber}
\newcommand{\nin}{\noindent}
\newcommand{\be}{\begin{equation}}
\newcommand{\ee}{\end{equation}}
\newcommand{\bel}[1]{\begin{equation}\label{#1}}
\newcommand{\bea}{\begin{eqnarray}}
\newcommand{\eea}{\end{eqnarray}}
\newcommand{\beal}[1]{\begin{eqnarray}\label{#1}}

\def\GG{\hat{\mathcal{G}}}
\def\Pm{\hbox to 6pt {$I\!\!\!\!P$}}
\def\F{\hbox to 10pt {\hfill \large $\cal F$} }

\begin{document}
\title{\bf  On the eikonal  unitarisation at high energies}
\author{ {\sc  O.V. Kancheli}\thanks{mail: kancheli@itep.ru} \\
     {\it  Institute for Theoretical and Experimental Physics, }  \\
     {\it  B. Cheremushinskaya 25, 117 259 Moscow, Russia. } }
\maketitle

\begin{abstract}
We consider approaches to the eikonal-like unitarization of
elastic amplitude and its generalizations in theories where
cross-sections grow with energy, and we discuss corresponding
mechanisms of the multiple exchange standing behind it. In
particular, we argue that in such theories the weight of the
$n$-fold exchanges can grow with $n$ much faster than for the
simplest Glauber eikonal.
\end{abstract}


\newpage
\setcounter{footnote}{0}
\section*{\bf 1. Introduction }

The eikonal unitarization  is a popular method that allows to
obtain the amplitude $F$, which satisfies some minimal s-channel
unitarity conditions from the ``non-unitary''  amplitude $A$. In
its simplest ( Glauber ) form, this method consists in the
replacing of the initial amplitude  $A \rightarrow F =  i (1 -
\exp{ (i A) } ) = -i \sum_n (iA)^n / n!$. This corresponds to a
summation of contributions from the  multiple exchanges, described
by the ``primary'' amplitude $A$, entering with equal weights.

Sometimes one can associate \cite{Glauber} with this approach a
picture of a fast particle moving along an almost straight line in
a target media with multiple scatterings, and when the full
S-matrix is the product of S-matrices of the individual
scattering. However, at high ultrarelativistic energies, the
longitudinal ``preparation length'' of the state of a fast
particle is much larger than the size of the interaction region.
So, this simple analogy breaks down, and all such multiple
exchanges must occur ``at the same time'', and they are needed to
take into account various effects, such as the effects of
screening and also describe new inelastic processes that are not
contained in the initial amplitude.

The mechanism of eikonalization can also be connected
\cite{eikonal} with the incrementing of a phase of the wave
function of a target particle, while passing through it the frozen
field of a fast particle.

In the Regge approach \cite{Gr-wc}, the eikonalization corresponds
to the summation of contributions of all multi-reggeon exchanges
(the non-enhanced Reggeon diagrams). Here the weights of the
n-reggeon exchange are given by the values of vertices $N_n$
describing the emission of $n$ reggeons by the external particle.
Their values are parameters which are defined outside the Regge
approach, and can, in principle, be calculated in the underlying
"microscopic" theory.~ And there is no particular reason to assume
that these weights have the simple Glauber form, especially when
the coupling is strong and nonperturbative as in QCD.

If the bare amplitude $A (s, t)$ grows with energy and becomes
large, then basically the high order terms of eikonal series
determine the dynamics of high energy processes and their values
should be chosen in agreement with the adequate theory. Usually,
however, especially in phenomenological calculations, these
eikonal weights are taken in a framework of simplest models,
without any specification.~ Here the most popular form is the
Glauber eikonal, or its slightly cured form - the quasi-eikonal
\cite{KaTer} expression for the weights of the n-fold exchange
chosen so as to take into account the contribution from
diffraction generation. For the same purpose, the matrix
generalization of the eikonal is also often used. But when the
pomeron contribution increases with the energy (as in QCD) and the
higher order eikonal terms become important such simple
corrections may not be sufficient.

In this article we review and discuss  various questions related
to the eiconalization  and, in particular, we argue that one
should expect a  much more rapid growth of weights of higher order
exchanges, up to the limiting behavior, when the eikonal series
may even begin formally diverge.

Our consideration is  carried out in the Regge approach, and in
this case there are not particularly significant the most of a
details of the specific field theory lying behind.

\vspace{3mm}

The paper is organized as follows.

In section ${\bf ~ 2}$ we recall and briefly overview different
approaches to eikonalization and also collect number of general
formulas for the multipomeron vertices $N_n$ and for
cross-sections in which these $N_n$ enter.

In section ${\bf ~ 3}$ we discuss some phenomenological models for
a generalized eikonal, that were used to describe the experimental
data.

In section ${\bf  ~ 4}$ we consider the structure of $N_n$
vertices in the parton model, in the QCD and in a number of other
approaches.

In section ${\bf  ~ 5}$ we estimate the behavior of $N_n$ vertices
in the $n \gg 1 $ limit.

Section ${\bf  ~ 6}$ contains a brief  Conclusion

\section*{\bf 2. ~~The general relations. Review.}
\vspace{1pt}

The eikonal unitarization of an elastic amplitude corresponds to
the summation of contributions from multiple exchanges, described
by some primary amplitude $A (s, t)$. In the Regge approach, this
procedure is reduced to summing of contributions of all
multireggeon exchanges (that is of all non-enhanced reggeon
diagrams). At a sufficiently high energy the amplitude $A (s, t)$
can be represented as a pomeron exchange or as a more complex set
of reggeon diagrams (Froissarons  in the limit of asymptotic
energies in QCD and in similar theories). This leads to the
expression for the unitarized (eikonalized) elastic amplitude
 \beal{ser}
 F(s,t) = \sum_{n=1}^{\infty} F_n(s,t)  ~~,~~~~~~~y = \ln s/m^2~~,
 ~~~~~~~~~~~~~~~~~~~~~ \\
 ~~F_n(s, t \simeq -k_{\bot}^2) = \frac{-i}{n n!}\int  N_n^2(k_{\bot i})
    \prod_{i=1}^n  \frac{d^2 k_{\bot i}}{(2\pi)^2} ~D(s,k_{\bot i})
    ~\delta^2(k_{\bot}-\sum k_{\bot i}) ~~,   \nn
 \eea
where the ``primary amplitude'' $A = F_1 = G (k_{\bot}) D (y,
k_{\bot}) G (k_{\bot})$ is taken in the  factorized form, and
where $D(s, k_ {\bot})$ - is the Green function of pomeron \Pm
~(or of some more generic object like Froissaron $\F$)~
 \footnote {For simplification of formulas we do not write
explicitly in (\ref {ser})  signature factors and suppose that all
nontrivial complexity of amplitude $A$ is included in $D$. We also
assume that both colliding  particles ($1 + 2 \rightarrow 1 + 2$)
are equal, so in all amplitudes enter factors of the type $N_n^2 $
and not $N_n^{(1)} N_n^{(2)} $}~.
 The vertex function $N_n (k_{\bot i})$, entering Eq.(\ref{ser}) and
describing the emission of $n$ pomerons with transverse momenta
$k_{\bot i}$ by the external particle, can be expressed
\cite{Gr-wc} via integrals of the product of $G$ vertices.
Properties of so defined $N_n$ will be considered in this paper.
The multiple exchange contribution of type (\ref{ser}) is always
contained in a full amplitude, regardless of the structure of the
``irreducible amplitude'' $D(s, k_i)$.

For a sufficiently large $s$ the direct interaction between
pomerons can become essential, and then the general contribution
to $F$, which takes into account all inter-pomeron interactions,
can be represented by diagrams Fig.1b,
\begin{figure}[h]
\begin{center}
\includegraphics[scale=0.70, keepaspectratio=true]{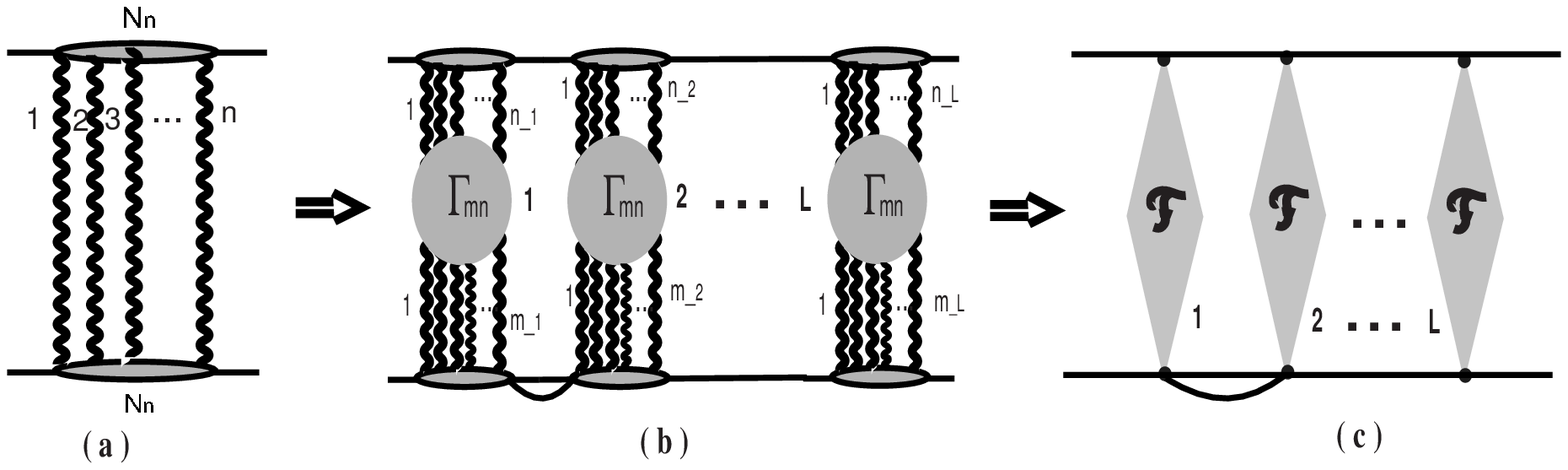}
\parbox{14cm}{    {\bf Fig.1~:}~~
 {\bf (a)} ~The non-enhanced reggeon diagram with general $N_n$
 vertices.
 ~~~{\bf (b)} The reggeon diagram with pomeron interactions in
  $\Gamma_{n m}$.
   ~~~{\bf (c)} The non-enhanced reggeon diagram with the Froisaron
   $\F$     exchange. This represents the limit of ({\bf b})
   at asymptotic energies.
} \label{Fig1}
\end{center}
\end{figure}
where the block $\Gamma_ {mn}$ corresponds to a general transition
amplitude of $n$ pomerons to $m$
  \footnote{ In this case, it may be useful to redefine the input
amplitude   from pomerons to some approximate froissarons  $A
\rightarrow \simeq \F$ }. ~Note in this regard that the average
rapidity values $\langle y_i \rangle$, on pomeron lines,
connecting vertices $ N_n $ and $\Gamma_ {mn}$, are not growing
when the full rapidity $Y = \ln s \rightarrow \infty $. Their
average values are determined by the magnitude of vertices of
inter-pomeron interactions (in particular by the 3 \Pm ~vertex
$r_{3P}$,... ) , so that $\langle y_i \rangle \sim g / r_{3P} \sim
$  the value of a ``free path in y of pomeron'' in the
multipomeron ``medium''.

At asymptotic energies the full amplitude $F$ can be represented
(Fig.1c) by the sum of non-enhanced reggeon diagrams with the
Froissaron ($\F$) exchange.  Note that such non-enhanced diagrams
must be included in the full amplitude $F$ with large weights,
because the elastic and diffraction generation cross-sections
(which are big at $s \rightarrow \infty $~;~~$\sigma_{elast}\simeq
\sigma_{tot} /2$ ) are connected with specific unitarity cuts of
the full amplitude.
\\

 Under rather general assumptions (in fact the same as for the Eq.
(\ref{ser})) the expressions for vertices  $N_n (k_i)$, entering
Eq.(\ref{ser}), can be written \cite{Gr-Mig} as an expansion over
on mass shell states of diffractive-like  beams
  \beal{decom}
  N_n(k_i) ~=~
  \sum_{\nu_1 ,\nu_2,.. \nu_n}\int
  ~G_{1 ~\nu_1}(P_{in},p_i^{1})~
  G_{\nu_1 \nu_2}(p_i^{(1)}, p_j^{(2)}) \cdots   ~~~~\nn \\
  \cdots  G_{\nu_{n-1} 1}(p_i^{(n-1)} , P_{out}) ~
 \prod_{i=1}^{n-1} d \Omega_{\nu_i}(p_i^{(1)})~~,~~~
  \eea
where ~ $G_{\nu_1 \nu_2} (p_i ^ {(1)}, p_j^{(2)}, k_{\bot})$ ~ is
the transition amplitude for a beam of $\nu_1$ particles with
momenta $p_i^{(1)}$ into a beam of $\nu_2$ particles with momenta
$p_j^{(2)}$, and with the emission of a ``pomeron'' with the
transverse momentum $ k_{\bot}$~
 \footnote{ In Eq.\ref{decom} it is also assumed that the
pomeron-particles amplitudes  $G_{\nu_1 \nu_2}$ entering $N_n$ duo
not contain contributions with the asymptotic behavior of leading
types (entering in  ($| A | $) , so that one can appropriately
deform all contours of integration in the masses of beams. That
is, the contributions from pomeron exchanges are not included in $
N_n $, (in particular, 3\Pm diagrams. ). But all contributions
from the secondary Reggeons are left in  $N_n$. }.
 ~In (\ref{decom}) the $d \Omega_{\nu} (p_i)$ is the element of
the $\nu$ particles phase-space volume. The expression (\ref
{decom}) contains a summation and an integration over all
kinematically allowed physical states of particles in beams
including full masses of beams.
\begin{figure}[h]
\begin{center}
\includegraphics[scale=0.55, keepaspectratio=true]{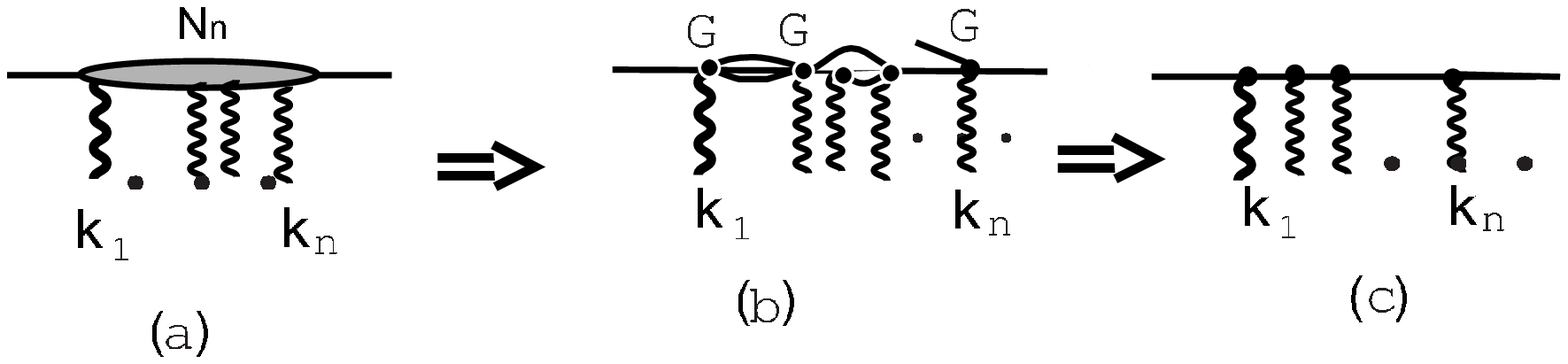}
\parbox{12cm}{    {\bf Fig.2~:}~~
 {\bf (a)} ~Diagram for $N_n$ vertex.
 ~~~{\bf (b)} Decomposition of vertex $N_n$ over diffraction beams.
 ~~~{\bf (c)} Pole approximation for  $N_n$ in the case of minimal
 (Glauber) eikonal.
} \label{Fig2}
\end{center}
\end{figure}
Note that all vertices $N_n (k_i)$ are real, while multiparticle
amplitudes $G_{\nu_1 \nu_2}$ are complex, and they also contain
disconnected contributions.

The Glauber eikonal form corresponds to a minimal contribution to
$N_n (k_i)$ from the single-particle state in beams. In this case,
all integrations in (\ref{decom}) disappear and the result is the
simple factored expression
 \bel{}
   N_n(k_i) ~=~ \prod_{i=1}^n g(k_i) ~,~~~~
 \ee
where $g (k) = G_{1 1} (p, p + k)$ - is the elastic pomeron
vertex. This leads to a great simplification and to the
exponential form of the elastic S-matrix in the impact parameter
representation
 \beal{exps}
   S(y,b) ~=~    S[v]   ~=~
   \sum_{n=0}^{\infty} \frac{(~i~v(y,b))^n}{n!} ~=~
    e^{i v(y,b) }~=~ 1+i F(y,b)~~,~~~~~\\
    v(y,b) ~=~ \int d^2 k_{\bot} e^{i b k_{\bot} }
    ~g^2(k_{\bot}) ~D(y,k_{\bot})~~,~~~~~y = \ln s/m^2   \nn
 \eea
For a large $y$ the value of Im $ v (y, b)$ can become large ~(
such as in the case of the supercritical regge pole $v \sim g^2
\vartheta (\Delta) \exp{ (\Delta y -b ^ 2/4 \alpha' y)} ~,~
\vartheta \simeq i ,~\Delta >0$  ),~ and then the expression
(\ref{exps}) corresponds to a picture of colliding black discs
with the Froissart asymptotic behavior.

The value of $|S(y, b)|^2$ is equal to the probability that both
fast particles will pass near one another ( at the impact
parameter $b$ ) without transition to the another states.
Therefore, the Glauber expression for $| S |^2 = e^{-2 Im (v)}$,
which follows from (\ref{exps}), can be interpreted in a simple
way. It is given by the Poisson probability that single particles
states pass through each other without interaction, and where the
value 2Im $v$ gives the average number of inelastic interactions
in such a process. The Glauber form of $S [v]$ given by
(\ref{exps}) corresponds to that all these 2~Im~$v(y,b)$
interactions are uncorrelated.

The expression (\ref{decom}) for $N_n (k_i)$ can be represented in
the symbolic operator form
 \bel{Noperat}
  N_n(k_i) ~=~ \langle  P_{in} | \GG (k_1)
     \GG (k_2) ... \GG (k_n) | P_{out} \rangle~ ~=~~~~~~~~~~~
  \ee
  \beal{Noperat2}
 =~ \sum_{\nu_1,...\nu_{n-1}}
   \langle  P_{in} | \GG (k_1)|\nu_1 \rangle~
   \langle \nu_1 |\GG (k_2)|  \nu_2   \rangle~
   \langle \nu_2 | \GG (k_3)|  \nu_3  \rangle   ...~~~~~~\\
~~~~~~~....~\langle \nu_{n-2} |
   \GG (k_{n-1})| \nu_{n-1} \rangle~
   \langle \nu_{n-1} | \GG (k_n)| P_{out}
   \rangle~~~,~~~~\nn~~~~~~~
 \eea
as the average of the product of non-local field operators $\GG
(k)$ describing the pomeron emission vertices $G_{\nu_1 \nu_2}
(k)$ between the initial and final state of the external particle,
and on the next step (\ref{Noperat2}) as the decomposition of this
product over the full system of physical states of beams $| \nu
\rangle $. ~If it is possible to redefine the bases for the beam
states  $| \nu \rangle$ so that ~ $ \GG (k) | \nu \rangle =
g_{\nu} (k) | \nu \rangle $ ~, ~that is to make all vertex
operators $ \GG (k) $ ~diagonal, then the expression
(\ref{Noperat2}) is simplified
 \bel{Ndiag}
   N_n(k_i) ~= \sum_{\nu} w(\nu)~
     \prod_1^n g_{\nu}(k_i)~~,~~~~~
    w(\nu) = \langle  P_{in} |\nu \rangle~
   \langle \nu |  P_{out} \rangle~,
 \ee
where  $w(\nu) $ is the probability to find the fast hadron in the
state $| ~\nu \rangle $. ~~After that the elastic $S(b,y)$ matrix
in an impact parameter representation can be represented in such a
form
 \beal{exps2}
  S(y,b) ~=~  \sum_{\nu_1 \nu_2} w(\nu_1) w(\nu_2)~
   \sum_{n=0}^{\infty} \frac{(~i~v_{\nu_1 \nu_2} (y,b))^n}{n!}
   ~=~~~~~\nn ~~~~~~~\\
   ~=~\sum_{\nu_1 \nu_2} w(\nu_1) w(\nu_2)~
   e^{i v_{\nu_1 \nu_2}(y,b) } ~,~~~~~~~~~~~~~~~~~ ~~~~~\\
    v_{\nu_1 \nu_2}(y,b) ~=~ \int d^2 k_{\bot} e^{i b k_{\bot} }
    ~g_{\nu_1}(k_{\bot}) ~D(y,k_{\bot})~ g_{\nu_2}(k_{\bot})~,~~~ \nn
 \eea
generalizing the Glauber eikonal expression (\ref{exps}).~ If we
assume that the dominant contribution to such a diagonalized
vertices (\ref{Ndiag}) is universal
 \bel{gnuk}
  g_{\nu}(k) ~=~ g(k) \lambda(\nu) + \tilde{g}(\nu, k)~,
 \ee
so that the non-factorisable part  $\tilde{g}(\nu, k)$ is
relatively small~
 \footnote{Such a factorization corresponds to a situation when
the properties of ``emitted pomerons'' do not depend on the number
of particles in beams and the value of $\lambda (\nu)$ is in fact
the number of ``average'' particle in the beam  $\nu$. This
factorized form of $g_{\nu}(k_\bot)$ can be expected only for a
small $ k_{\bot}$, since for large $k_{\bot}$ the contribution of
states $ | ~\nu \rangle $ of a small transverse size  $ \sim
k_{\bot}^{-1}$ can be most essential. ~ However, for large
$k_{\bot}$ the contributions of higher exchanges are not so
significant, and for small $k_{\bot}$ the value of $ \beta_n $ is
roughly proportional to the number of particles in beams, because
the individual pomerons are emitted mainly by the different
particles in beams.},
 then we come to the representation
 \bel{geik}
   N_n(k_i) ~\simeq~    \beta_n~ \prod_1^n g(k_i)~~,~~~~~~
   \beta_n  ~=~   \sum_{\nu} w(\nu)~(\lambda(\nu))^n ~.
 \ee
The form (\ref {geik}) for $ N_n $ (which is somewhat simplified
in comparison with (\ref{exps2}) ) is such a generalization of the
Glauber eikonal which we will mainly consider below~
 \footnote{Discussion of a more general case is in
 Section 2c and in Section 3}.
 Turning to the impact parameters as in (\ref{exps}), we obtain
the generalized eikonal series
 \bel{sexp}
   S(y,b) = \sum_{n=0}^{\infty}
   \frac{\beta_n^2}{n!}~
   (~i~v(y,b))^n~,~~
 \ee
where the real weights $\beta_n \ge 1$ can be almost arbitrary. It
is useful to present the series (\ref{sexp}) in a compact form,
resembling the Glauber exponential.
One can do this in a number of ways. Here we use  one of them.
Substituting in (\ref{sexp}) coefficients $\beta_n$ in the form
 \bel{bet}
    \beta_n = \int_0^{\infty} d \tau~\tau^n~\varphi(\tau)~,
 \ee
we can write $S [v]$ as a superposition of Glauber like eikonal
contributions
 \beal{smat}
   S(y,b)\equiv S[v] =
   \int_0^{\infty} d\tau \rho(\tau)
    ~e^{i\tau v(y,b) }~,~~~
   \rho(\tau) = \int_0^{\infty}
    \frac{d\tau_1}{\tau_1}
   ~\varphi(\tau_1)~\varphi(\tau/\tau_1)~,
 \eea
entering with weights $\rho (\tau)$ ,~ where $\tau_i g$ act as the
effective pomeron vertex for the $|~\tau >$ state.  The
normalization condition for $S[v]$ and $w (\nu)$ leads to
conditions $\beta_0 = \beta_1 = 1$, and this implies relations
 \bel{norm}
   \int_0^{\infty} d\tau \rho(\tau) =~
   \int_0^{\infty} d\tau ~\tau \rho(\tau) = 1~~.
 \ee
Note also the convexity condition
$$
 \beta_n \beta_m \le \beta_{n+m}~~,
$$
following from the inequality
$$
 N_n(k_1,... k_n)~N_m(q_1,...q_m) \le
    N_{n+m}(k_1,... k_n,q_1,...q_m)~,
$$
which leads to the following relation for the $S [ v ]$-matrix as
a function of the amplitude $v$
$$
 ~~~S^2[x] ~\le~ S[2 x]~~~,~~~~for ~ Re ~x = 0~,~ Im~x < 0 ,
$$
and where the equality is only for the Glauber eikonal case
(\ref{exps}).

The Glauber case corresponds to the simple density
  \be
    \rho(\tau) ~=~ \delta (\tau - 1) ~,
 \ee
and for the matrix model with $L$-states the spectral function is
 \be
   \rho(\tau) ~=~ \sum_{n=1}^L
    c_n ~\delta (\tau - \tau_n)~,~~~
   \tau_n ,~ c_n \ge 0,~~\sum_{n=1}^L c_n =\sum_{n=1}^L \tau_n c_n = 1 .
 \ee
In addition we also list a number of general relations for
cross-sections at given impact parameter
 \footnote{In the Eq. (\ref{totcs} - \ref{npcut}) all cross-sections
are given at certain \underline{fixed} values of the  impact
parameter and are dimensionless. The corresponding ``usual''
quantities are given by $\int d^2 b ~\sigma{...} (y,b)$ }
 , which are expressed in terms of the function $S [ v (y, b) ]$
and which are valid for an arbitrary spectral density
$\rho(\tau)$~:

\nin the total cross-section
 \bel{totcs}
\sigma_{tot}(y,b) = 2(1 - Re~S[v])~~~,~~~~~~~~~~~~~~~~~~~~~~~~~  \\
 \ee
the elastic cross-section
 \be
\sigma_{el}(y,b) = |~1 - S([v]~|^2~~~~,~~~~~~~~~~~~~~~~~~~~~~~~~~~~~\\
 \ee
the total inelastic cross-section
 \be
 \sigma_{in}(y,b) = \sigma_{tot}
-  \sigma_{el}  =
         1 - |S[v]|^2~~~,~~~~~~~~~~~~~~~~~~  \\
 \ee
the pionization  cross-section - corresponding to processes when
at least one pomeron is s-cut
 \be
\sigma_{\pi}(y,b) = 1 - S[2 i Im~v]~~,~~~~~~~~~~~~~~~~~~~  \\
 \ee
the total cross-section of diffraction generation (single -
$\sigma_{sd}$ and double - $\sigma_{dd}$)
 \bel{difgen}
   \sigma_{dif}(y,b) =  \sigma_{in} - \sigma_{\pi} =
    2 \sigma_{d} + \sigma_{dd} =
      S[2 i Im~v] -|~S[v]~|~^2~~.~~~~~~~~~
 \ee
The expression
 \bel{npcut}
 \sigma_n(y,b) = \int_0^{\infty} d\tau \rho(\tau)
 ~\frac{(2\tau Im~v)^n}{n!} ~e^{-2 \tau Im~v}  ~~,~~
 \sigma_{\pi} = \sum_{n=1}^{\infty} ~\sigma_n~~
 \ee
gives the cross-section for $n$ soft ``multiperipheral'' beams
(these are the contributions of diagrams with $n$ cut pomerons and
of the arbitrary number of uncut pomeron lines) as a superposition
of Poisson distributions.

\nin The quantity
 \bel{njet}
   \overline{n(y,b)} =
   \frac{\sum_n n~\sigma_n (y,b)}{\sum_n \sigma_n}
   = 2 Im ~v(y,b) ~,
 \ee
as it follows from (\ref{npcut}) and (\ref{norm}), gives the
average multiplicity of cut pomerons as a function of $(y, b)$.~
It is interesting that this quantity does not depend on the form
of $\rho (\tau)$,~ i.e. on values of $\beta_n$.

\subsection*{ \it{2a~~
Field theory approach to eikonal } }

In the field theory approach the eikonal S-matrix is defined
\cite{eikonal} by the operator expression $\hat {S} \simeq T
\exp{(i \int L_{int})} ~ $, where the T-ordered integration is
taken along  straight paths of fast particles. It is convenient to
describe this scattering in the laboratory frame of one of the
colliding particles. The field of the fast charged particle with
energy $E$ is concentrated in the longitudinally compressed thin
($ \sim 1 / E $) sheet  and it is frozen for the time of
collision, and therefore it can actually be regarded as a
classical perturbation while this field disc passes through the
target. Under these conditions, the elastic S-matrix is given by
the expression ~$\langle out | \hat {S} | in \rangle = e^{i
\delta}$, where $\delta (E,b) = A (E,b)$ is a phase gained by the
wave function of a ``target particle'' when the fast disk-particle
passes through it. This directly leads to the Glauber expression
(\ref{exps}) for~$S[A]$.

However, such a simple expression for the $S$ does not account the
possibility that the incoming particle can fly up to the target
being in a different ``valence'' states, and therefore carry a
different ``classical''  fields. Moreover, there can be
significant fluctuations of field values on different energy
scales ( which corresponds to the parton cascading ). This last
one, in particular, leads to a longitudinal spreading ( up to
$\sim 1 / m$ ) of the field incident on the target particle.

Some of these effects can be accounted for by summing over all
possible valent states (classical fields) of a fast particle
approaching the interaction region:
  \bel{eik2}
    \langle out|\hat{S}|in\rangle ~=~ \sum_{\alpha ,
    \beta} \langle  out|U(\infty,t_0)|\alpha \rangle
    ~\langle \alpha|U(t_0,-t_0)|\beta
    \rangle ~\langle\beta|U(-t_0,-\infty)|in>~,
  \ee
where $| \alpha \rangle, ~ | \beta \rangle $ are the ``parton''
states of fast particles,  involved in the interaction, and where
$t_0 \gg 1 / m$, because the preparation time of such states is
usually much larger than the interaction time. ~Leaving here only
diagonal states, which correspond to the approximation that one
neglects possible changes of the fast particle field in the
interaction time interval $(-t_0 ~ \div t_0)$, we obtain
 \beal{eik3}
  \langle out|\hat{S}|in> ~=~
   \sum_{\alpha} \langle out|U(\infty,t_0)|\alpha\rangle
   \langle\alpha|U(t_0,-t_0)|\alpha \rangle
   \langle\alpha|U(-t_0,-\infty)| in \rangle ~\simeq~
   \nn\\
   ~\simeq~ \sum_{\alpha} |\langle in|U(-\infty,t_0)|
     \alpha\rangle|^2
   ~\langle\alpha|U(-t_0,t_0)|\alpha\rangle ~\simeq~
     \sum_{\alpha} w_{\alpha}
    e^{i\delta_{\alpha}}~~~,~~~~~~~
 \eea
where quantity
$$
w_{\alpha} ~=~ |\langle in|U(-\infty,t_0)|\alpha\rangle|^2
$$
gives the probability that the fast incoming particle is in the
state $| \alpha \rangle $ directly before the interaction with a
target, and the $\exp{( i \delta_{\alpha})} = ~ \langle \alpha | U
(-t_0, t_0) | \alpha \rangle $ - is a factor acquired by the wave
function of the target while passing through the field of the fast
particle in the state $| \alpha \rangle$. ~The expression (\ref
{eik3}) for S has the same structure as the spectral
representation (\ref {smat}).

\subsection*{ {\it 2b~~ The behavior of  $S[v]$   at large $|v|$ }}

The behavior of $ S [v] $ at  $| v (y, b) | \gg 1$ corresponds to
the high energy limit $ y \gg 1 $, because usually $v(y,b) \sim
f(\tilde{y}, b) \exp{(\Delta \tilde{y})}, ~ ~ \tilde{y} \simeq y
-i \pi / 2 ,~  \Delta > 0$. As follows from (\ref{smat}) the
asymptotic behavior of $S [v \rightarrow \infty]$ is determined by
the spectrum $\rho (\tau)$ at small values of $\tau$. If $\rho
(\tau) = 0$ for $\tau < \tau_0$ then $S [v] \sim \exp {(- \tau_0 |
v |)}$. This occurs, for example, in finite matrix models for
vertices $N_n $.

The general case, when the spectrum $\rho (\tau)$ extends to $\tau
= 0 $, is more interesting. If $\varphi (\tau = 0) \ne 0$, then,
as follows from (\ref{bet}), ~$\rho (\tau) \sim \ln 1 / \tau $ for
$ \tau \sim  0$~,~ and we obtain from (\ref {smat}) the asymptotic
expression for $S [v]$ at $ | v | \gg 1 $
 \bel{sv}
   S[v] ~\simeq~   c_0 \frac{\ln{\hat{v}}}{\hat{v}} ~+~
   \frac{c_1}{\hat{v}} ~+~ \frac{c_2}{\hat{v}^2} ~+...
    ~,~~~~~~\hat{v} = -i\vartheta |v| \simeq Im v ~.
 \ee
Similarly, if $\varphi (\tau \ll 1) \sim \tau^\lambda, ~ \lambda >
-1 $, we get $ S [v] ~ \sim 1 / \hat {v}^{1 + \lambda} $. It is
also natural to suppose that $\rho (\tau)$ has no singular
contributions of the type $\delta (\tau)$, because they correspond
to such components of Fock state of a fast particle that do not
interact with the target and cause the asymptotic $S[v \rightarrow
\infty] \rightarrow const$.

Note that the expression for the eikonal series (\ref{sexp}) can
be represented
  \footnote {Similar method was used \cite{Cardy} to compute the
Froissart asymptotic, arising from the summing of  contributions
of all reggeon diagrams in the case of large $v$.}
 in the form
 \bel{card}
   S[v] ~=~ - \int_{c - i \infty}^{c + i \infty}
   \frac{d n}{\sin \pi n} \Gamma (1-n) v^n \beta^2_n~,
 \ee
when using the continuation of  $ \beta_n^2 $ to  non-integer $n$
~. The same analytic continuation of $\beta_n$ defines the
behavior of the spectrum
 \bel{ft}
 \varphi(\tau) ~=~ - \int_{c - i \infty}^{c + i \infty}
 \frac{d n}{2 \pi i ~\tau^{n+1}}  ~\beta_n~~.
 \ee
The distribution of singularities of the function $\beta_n$ is
connected
 \footnote {It follows from the definition (\ref{bet}) for $
\varphi (\tau) $  that $\beta_n$ is analytic in Re $n> -1$ for
sufficiently fast decreasing  spectrum at $ \tau \rightarrow
\infty $ ;~ (decreasing as $\ln \varphi (\tau) < -c \tau $ ~, or
faster). }
 with the properties of spectrum $\varphi (\tau)$. As
follows from (\ref{card}) the behavior (\ref{sv}) is determined by
the rightmost singularity of $\beta^2 (n) $ for $n = -1 $. The
expression (\ref{exps}) describing the Glauber eikonal corresponds
to $\beta_n^2$ regular for all finite $n$. A more general case,
when the spectral density $\varphi (\tau \ll 1) \sim \tau^\lambda,
~ \lambda> -1 $ , corresponds to the rightmost singularly $\beta_n
$ at $ n = -1 - \lambda$.

\subsection*{ \it{2c~~
Eikonal  with many different virtuality scales } }

As was noted above the factorization of the type (\ref{gnuk}) for
vertices of $G_{\nu \nu} (k_{\bot})$ should not be expected in the
general case, and the dependence of amplitudes  $V_{\nu_1 \nu_2}
(k_{\bot})$ on the pomeron momentum $k_{\bot}$ may be various for
different states $\nu_i$. This is well illustrated by a simple
example of a model, when $V_{\nu_1 \nu_2}$ is the contribution of
a certain Regge pole, and when the dependence of vertices from $r$
is taken (for simplicity) in the exponential form
$$
 V_{\nu_1 \nu_2} ~=~
    \frac{g_{\nu_1} ~g_{\nu_2}}{~4 \pi R^2_{\nu_1 \nu_2}}~
  \exp{ \Big( \Delta \tilde{y}  -b^2 / R^2_{\nu_1 \nu_2} \Big)}~,
$$
where $R^2_{\nu_1 \nu_2} = R^2_{\nu_1} + R^2_{\nu_2} + 4 \tilde{y}
\alpha'_{\nu_1 \nu_2}  ,~ \tilde{y} = y - i\Delta/2$. ~ Then for
soft beams with a large mass the value of $R^2_{\nu_1 \nu_2}$ can
be large. And for those components of $\nu_i$, to which more hard
components of pomeron are attached, the value of $R^2_{\nu_1
\nu_2}$ can be relatively small.

One can expect that in theories (like QCD), where the effective
coupling decrease with the growth of virtuality, the spectrum of
pomeron states would be discrete. Under these conditions, the
generalized eikonal series for the diagonalized amplitude (as in
(\ref{sexp})) can be approximately written as a sum of
contributions arising at the different scales of virtuality
 \bel{mpol}
   F(y,b) ~=~ \sum_{\mu} \sum_{n = 1}^{\infty} \beta^2_n (\mu)
    \frac{(~i v_{\mu})^n }{n! }~~,~~~~
    \sum_{\mu} \beta^2_1 (\mu) ~=~1~,
  \ee
where $v_{\mu} (y, b)$ - is the primary scattering amplitude, on
the $\mu$-th scale of virtuality. It is also assumed that in the
expression (\ref{mpol}) the virtuality of the beam state does not
change
 \footnote{These virtualities are mainly related to transverse
dimensions of colliding states and they do not have enough time to
change during the collision}
 at  vertices of $\GG (\mu_i)$ ~. This corresponds to a
relative independence of scattering processes on different
virtuality scales.

An example of these amplitudes can be the simple Regge
contributions
  \bel{npom}
  v_{\mu}(y,b)) \simeq v_{\mu}^{(0)} \exp{(\Delta_{\mu} y -b^2/4y
  r_{\mu}^2)}  ~~,~~~~ \mu = 1,2,3,...
  \ee
coming from the chain of poles \cite{multpom} with intercepts
$\Delta_{\mu} \sim 1 / \mu$ and with the increasing virtualities,
corresponding to the average transverse dimensions $r_{\mu}^2 ~
\sim \exp{(- c\mu)} $. The amplitudes (\ref{npom}) can be combined
to give the BFKL type Pomeron which the QCD coupling running with
scale, and for with it is also possibile to include the main
non-perturbative effects, adjusting the value of first intercepts
$\Delta_{\mu}$.

A similar to (\ref{mpol}) type expression for $F$ can be obtained
\cite{BrowerStrTan} if we consider the high energy scattering
process in 5-dimension and connect the average pomeron virtuality
with the radial AdS coordinate.

The total $S(y, b)$ matrix corresponding to (\ref{mpol}) can be
written as the sum of $S_{\mu}$-matrices related to different
scales of virtuality
 \bel{ssup}
   S(y,b) ~=~ \sum_{\mu =1}^{\infty} ~S_{\mu}(y,b)~,~
 \ee
$$
  S_{\mu}(y,b) ~=~ \sum_{n=0}^{\infty} (\beta_n^{(\mu)})^2
  \frac{(~i ~v_{\mu}(y,b))^n}{n!}~,~~\sum_{\mu} (\beta_0^{(\mu)})^2 =1~.
$$
The same expression for $S$ in the exponential form, similar to
(\ref{smat}), is
 \bel{ssup2}
   S(y,b) ~=~ \sum_{\mu}   \int_0^{\infty}
     d\tau \rho(\tau,\mu)
     ~\exp { \big ( i\tau v_{\mu}(y,b) \big ) }~,
 \ee
where the spectral density $\rho(\tau, \mu)$ depends now also on
the virtuality $\mu$.

Expressions (\ref{ssup} - \ref{ssup2}) correspond to a simple
physical picture that the total amplitude for the particles to
pass without interaction at the impact parameter $b$ is given by
the superposition of amplitudes to pass without interaction in
different virtual states, because these states at a very high
energy are prepared during times which are much longer than the
time of the collision.

In theories with the $\Delta_{\mu} \sim 1 / \mu$ pomeron sequence
when $y$ increases the gradual inclusion of higher $S_k$ matrices
with a more and more large virtuality take place. The
corresponding picture of the Froissart limit will be such that the
parton structure of a fast particle can be represented as a system
of nested more and more hard disks expanding with the increasing
of energy.

 \vspace{5mm}

\section*{\bf 3.~~
Some phenomenological models for the generalized eikonal }

In this section we briefly review a number of popular models for
vertices $N_n \equiv g^n \beta_n$, generalizing the simple Glauber
eikonal. Such models were used to describe the experimental data
on hadron elastic and inelastic processes at high energies, and so
they reflect many of properties of ``correct'' $N_n$ vertices,
which unfortunately can not be calculated using the existing now
QCD methods.

\vspace{3mm}

1) The simplest generalization of the Glauber eikonal is the
quasieikonal model ~ \cite {KaTer}. ~ Here eikonal coefficients
are chosen in the form
 \bel{qeik}
   \beta_n^2 = C^{n-1} + \delta_{n 0} (1 - C^{-1})~,
 \ee
where the parameter $C$ represents a contribution of the average
beam in  (\ref {geik}) and it is directly related to the
cross-section of diffraction generation $\sigma_{dif} /
\sigma_{el} = C-1$. From a comparison with various experimental
data at not too high energies it follows that $C \simeq 1.2 $. The
spectral density has the form
$$
  \rho(\tau) = C^{-1} \delta (\tau - C) +
                   (1 - C^{-1}) \delta (\tau)~.
$$
In this case $S(b, y) = 1 + C^{-1} (\exp {\big (i C v (b, y)
\big)} - 1) $. In fact, this model assumes the existence of two
states of a fast hadron - the single-particle component
interacting with the vertex $g$, and of a naked component entering
in the spectrum at $\tau = 0 $, and which at all does not
participate in the interaction. This, in particular, leads to a
nonanalyticity of $\beta_n$ at $n = 0$. Probably, such
non-interacting states corresponding to  $ \delta (\tau) $ type
contribution to spectrum, can not occur
 in reasonable field theories. The simplest corrected expression
$$
 \beta_n^2 =  \frac{n (2-C) + C}{1+n}~C^{n-1}
$$
has approximately the same properties as (\ref{qeik}) but does not
violate relations (\ref {norm}) and at large $n$ give $ \beta_n^2
\simeq C^{n-1} (2-C) $. \vspace{1mm}

Quite indicative is the one-parameter model, with extremely rapid
growth of eikonal coefficients
 \bel{fac}
   \beta_n^2 = \frac{\Gamma(n+a)}{a^n ~\Gamma(a)}~,
 \ee
and where the eikonal series converge rather poorly for amplitudes
$v$ growing with energy. The corresponding  spectral density is
given by
$$
 \rho(\tau) = \frac{a^a}{\Gamma(a)} ~\tau^{a-1} ~e^{-a \tau}~,
$$
and the $S(b,y)$-matrix is
 \bel{pred}
    S[v] ~=~ \Big(1 - i v/a \Big)^{-a}~.
 \ee
The representation (\ref{pred}) looks particularly simple in the
extreme case when $a = 1$~.~~ All expressions  (\ref{totcs} -
\ref{npcut}) for cross sections are also strongly simplified in
this case:
$$
 \sigma_{tot}(y,b) = \frac{2 v_2}{1+ v_2}~;~~
 \sigma_{in} = \frac{2 v_2}{1+ 2v_2}~;~~~~
  \sigma_{dif} = \frac{| v |^2}{(1+ 2v_2)( 1+ 2v_2 + | v |^2 )} ~;
$$
$$
  \sigma_n(y,b) =  \frac{1}{1 + 2 v_2}~\Big(
     \frac{2 v_2}{1+ 2 v_2} \Big)^n ~;~~~~~~
     v_2 = Im v~;~~~
$$
The form of $S[v]$ similar to (\ref{pred}) was discussed many
times in the literature. In the next sections we present some
dynamical arguments that the behavior of this type can be
naturally realized for theories (like QCD) with a growth with the
energy cross-sections.

For the not smooth at $\tau = 0$ spectrum $\rho (\tau) = c_1
\delta (\tau) + \rho_1 (\tau) $, where the dependence of $\beta_n$
on $n$ is nonanalytic, we come to a non-decreasing with $| v|
\rightarrow \infty$ contribution to $S [v]$. An example of such a
behavior, which is  close to the (\ref{fac}), is the one-parameter
model
 \bel{fac2}
   \beta_n^2 = c^{n-1} n! + \delta_{n 0} (1 - c^{-1})~~,~~~~
  \rho(\tau) = c^{-2} e^{- \tau/c} + \delta (\tau) (1 - c^{-1})~,
 \ee
for which
$$
 S[v] ~=~  \frac{1 +i v (1 - c)}{1 - i v c}
$$
If $c = 1$, then this expression obviously coincides with
(\ref{pred}) for $a = 1$
 \footnote {~~If $ c = 1/2 $, this expression leads \cite{Tyurin}
to an elegant form ~ $S = (1 + iv / 2) /(1-iv / 2) $. For this
case $|S[v]| \leq 1$,  because  Im $ v \geq 0 $. For $|v|
\rightarrow \infty$ we have $S[v] \rightarrow -1$~. }

\vspace{3mm}

2) The poly-eikonal model \cite{polieik} is defined by relations
 \bel{poleik}
  \beta_n =  \prod_{k=1}^{n-1} \xi(n,k)~,~~~
  \xi(n,k) \simeq (1 +\frac{\lambda}{k + a})^{n-k}~~.
 \ee
It is a simple two-parameter generalization of the Glauber
eikonal, where the average multiplicities of particles in beams
grow with $n$, and where each factor $\xi (n, k)$ corresponds to a
beam with $\sim n-k$ particles.  The expression (\ref{poleik})
effectively takes into account the non-planar structure of $N_n$
and also includes the disconnected contributions to $G_{\nu_1
\nu_2}$ vertices. At large $n$ from (\ref{poleik}) one can obtain
that $\beta_n ^ 2 \sim (n !)^{\lambda}$. The coefficients
$\beta_n^2$ can be analytically continued in $n$ so that the only
singularity at finite n is the pole at $n = -1-a$.

\vspace{3mm}

3) The matrix models for $N_n$.  Here the spectrum of beams is
approximated by the discreet sequence of states $| \nu \rangle $
in Eq. (\ref{Noperat}). ~That is all multiparticle beams are
replaced by $L$ resonances, located on the mass shell. In this
case vertices $G_{\nu_1 \nu_2} $ describing the transition between
states $| \nu_i \rangle $, ~$ i = 1,2, ..., L$ are the real
symmetric $L \bigotimes L$ matrices. Choosing the value of $L$
large enough and tuning parameters of these matrices, one can
obviously get a description of $N_n$ vertices with any degree of
accuracy. Then the diagonalization can always be done explicitly
by the orthogonal transformation of basis $| \nu_i \rangle$, and
we get
 \beal{matrs}
 S(y,b) ~=~ \sum_{n=0}^{\infty}~\frac{1}{n!}~
 \sum_{\nu_1 \nu_2}  w_{\nu_1} w_{\nu_2}
 \Big(- i V_{\nu_1 \nu_2} \Big)^n ~~=  \\
  =~\sum_{\nu_1 ,\nu_2}^L
  w_{\nu_1} w_{\nu_2}~e^{- i V_{\nu_1 \nu_2} }~~~~~~~,~
 \eea
where $w_{\nu}$ is the probability to find a component  $ | \nu
\rangle$ in the $| in \rangle$ state, and
  \bel{aaa}
     V_{\nu_1 \nu_2} ~=~ \int d^2 k_{\bot} e^{i b k_{\bot} }
        ~G_{\nu_1 \nu_1}(k_{\bot}) ~D(y,k_{\bot})
        ~G_{\nu_2 \nu_2}(k_{\bot})~~~~
  \ee
are diagonalized  amplitudes. It is clear that these expressions
also follow directly from the general representation (\ref{exps2})
for $S$ if we assume that the spectrum $w(\nu )$ is discrete.

If we suppose additionally the factorization for vertices $G_{\nu
\nu} (k)$ of the type (\ref{gnuk}) then
$$
V_{\nu_1 \nu_2}(b,y) \simeq \beta_{\nu_1}\beta_{\nu_2}~v(b,y)~.
$$
In this case
$$
\varphi (\tau) =\sum_{i=1}^L c_i \delta(\tau -\tau_i)~;~~~ S(y,b)
~=~ \sum_{i,j=1}^L ~c_i c_j e^{ i \tau_i \tau_j~v(b,y) }~~.
$$

\section*{\bf 4.~~
$ N_n $ vertices in the parton model and scattering of
constituents}
\vspace{2mm}

\subsection*{ \it{4a~~
Constituent scattering } }

It is essential that only the non-planar Feynman diagrams give the
contribution to the multipomeron vertices $N_n$. This is usually
interpreted so as if all parton chains (~corresponding to pomeron
amplitudes $v$~) must exist simultaneously in the state of fast
particles and these chains are attached to the different
``valence'' partons, at that colliding particles ``split''. In
such an approach it is possible not to specify all the properties
of these primary partons and take into account only the fact that
each external hadron is some superposition $\sum a_m | m \rangle$
of states $| m \rangle$ containing $m$ such partons with
probabilities $ w_m = | a_m |^2$.

Then in the reggeon diagrams with the multi pomeron exchange one
can use some pomeron-like amplitude $\lambda^2 v (y, b)$
describing the elastic scattering of valent parton. Here $\lambda
= g_0 / g$ is the ratio of the parton $g_0$ to hadron $g$
vertices, describing the pomeron emission. Under these
assumptions, the amplitude of elastic hadron scattering can be
written as
 \bel{pamo}
  F^{(a,b)}[~v(y,b)~] ~=~
    \sum_{m1,~ m2}^{\infty} ~w_{m1}^{(a)}
   ~w_{m2}^{(b)} ~f_{m1,m2} [v]~,
 \ee
where $f_{m1,m2}[v]$ is the ellastic scattering amplitude
 \footnote{The amplitude $\lambda^2 v(y, b)$ entering this
expression include integrations over the transverse coordinates of
partons, with some vertex factors with respect to the center of
parton beams.}
 of partons in states $| m1 \rangle $ and $ | m2 \rangle $. It can
be  expressed in terms of the  $\lambda^2 v$  from a purely
combinatorial consideration, as the sum over all possible
rescatterings of partons
 \bel{apu}
   f_{m1,m2} [v] ~=~ -
   \sum_{n=1}^{\infty}~n! ~C_{m1}^n
  ~C_{m2}^n ~(-\lambda^2 v(y,b))^n~~,
 \ee
Here each binomial coefficient $ C_{m}^n $ corresponds to a
selection of $n$ partons participating at  scattering, from the
available in states $| m1 \rangle \bigotimes | m2 \rangle $, and
the factor $n!$ responds to different possible versions of
interaction of these partons.

Substituting the amplitude (\ref{apu}) to (\ref{pamo}) and
comparing $F^{(a,b)}$  with the generalized eikonal form
(\ref{sexp}) we obtain such an expression
 \bel{betw}
  \beta_n ~=~ n!~\lambda^n ~\sum_{m=n}^{\infty} C_m^n ~w_m~,
 \ee
connecting vertices $N_n = g^n \beta_n$ with weights of primary
parton states. Inverting these relations we obtain expressions for
parton probabilities $w_m$ in terms of eikonal factors $\beta_n$ :
 \bel{wbet}
  w_m ~=~ \frac{1}{m!~\lambda^m} ~\sum_{n=0}^{\infty}
     \frac{ (-\lambda^{-1})^n   }{n!} ~ \beta_{n+m}~~.
 \ee
Substituting in (\ref{wbet}) expressions for $ \beta_{n + m} $ in
the spectral form (\ref{bet}) we come to the representation of
parton probabilities
 \bel{wbet2}
    w_m ~=~ \int_0^{\infty} ~d \tau   ~P_m (\tau /\lambda)
        ~\varphi  (  \tau ) ~,
        ~~~P_m (\tau ) = \frac{1}{m !} \tau^m e^{-\tau}
 \ee
as a superposition of the Poisson distributions with the average
$\langle m \rangle = \tau $, and with weights $\varphi (\tau)$
defined by the spectrum of diagonal states of diffractive beams.
It is evident from (\ref{wbet2}) that Glauber eikonal corresponds
to a simple Poisson distribution of number of valent constituents
$w_m$ in the fast hadron with the average $\langle m \rangle =
\lambda^{-1}$.

As follows from (\ref{wbet2}) the behavior of $w_m$ at large $m$
is related to the properties of spectrum $\varphi (\tau)$ at $\tau
\gg 1 $. ~If the spectrum $\varphi (\tau)$ is bounded from above,
so that $\varphi (\tau) = 0$ for $\tau  > \tau_{max}$, as it is in
the case of finite matrix models, then the tail of $w_m$
distribution has again the Poisson form $w_m \sim \tau^m_{max} /
m!$. ~For a unlimited from above spectrum $\varphi (\tau)$ the
decrease of $w_m$ can be slower.  So for the Gaussian tail, when
$\ln \varphi (\tau) \sim - \tau^2$, the asymptotics is flatter
$w_m \sim c^m / \sqrt {m!}$ ~And for the case $ \ln \varphi (\tau)
\sim - \tau $, the tail takes power form $w_m \sim c^m, ~ ~ c <1$.

\nin For a large $n$ the general approximate relation
 \bel{basim}
  \beta_n ~\simeq~ n!~\lambda^n ~w_n
 \ee
follows from (\ref{betw}). Therefore, for Glauber eikonal and for
the finite matrix models when $w_m \sim c^m / m!$~,~ we have that
$ \beta_n \sim \lambda_1^n $ ~, ~ and  the Gaussian spectrum for $
\varphi (\tau) $ leads to $ \beta_n^2 \sim n! $

\subsection*{\it{ 4b~~
Parton picture in Fock representation and eikonal } }

The approach to high energy scattering based on the parton picture
can give a somewhat different view on the role of the multiple
scattering included to the eikonalization, since in this case all
restrictions, following from the $s$-unitary, are more simple and
clear.

Consider the collision of a fast particle with high energy $E = m
e^Y$ in the laboratory frame of another particle, when the elastic
amplitude $v$ is given by a reggeon (Pomeron) ladder with $\Delta>
0$. In the corresponding parton description the main components of
Fock wave function $| \Psi (Y) \rangle$ of a fast particle
correspond to states, created by the parton cascade, which
contains $\sim | v | \sim v_0 \exp{\Delta Y}$ ~low-energy partons.
During the collision mainly these low-energy partons interact with
the target particle. This picture is approximately true for all
field theories with $\Delta> 0$, including QCD.

If the energy $E$ is not particularly large, or when $ g_P^2 $ is
small, so that $| v | \ll 1 $, ~ then the multiple rescattering
(eikonal correction) are not so important, although the elastic
scattering and diffraction generation are contained in the $N_2^2$
eikonal term.

At $ |v| \sim $ 1 the eikonal corrections are of the same order as
the first term (single exchange).

Here the  main ``destination'' of rescatterings (corresponding to
non-enhanced reggeon diagrams) is firstly to take into account
various screening in the interaction of target particle with
parton ``medium'' incoming with the fast particle. They also
describe various new processes arising from multiple interactions
and the diffraction.

If the energy is so large that $| v | \gg 1$, then in the eikonal
series many terms  ($\sim | v |$~) are essential. But for such a
large $y$ the parton merging becomes significant, and as a result
the partons sequential saturation takes place, on scales with the
virtuality increasing with y.  And later the parton system
transits to the Froissart phase when the saturated parton disk
expands with y , and gradually  becomes more and more black as the
mean virtuality of partons grows.

To better understand how the S-unitarity is restricted by the
eikonalization it is helpful to consider the special case when the
colliding energy is very large, so that $| v | \gg  1$ , but the
interaction between pomerons is switched off.  This means that one
must not include the contributions of enhanced reggeon diagrams
 \footnote {This mode is almost realized in QCD in a certain range
energy, since the  $3 \Pm $ vertex is relatively small -  $r (3
\Pm) / g (h h \Pm) \ll~1$}.
 On the parton language this regime corresponds to a model when the
partons in cascades are not glued one with another. The parton
disk of a fast particle is on average very tightly packed, and the
density of low-energy partons infinitely grows with energy. In
this case the probability $W$ for the target particle to tunnel
through such fast dense parton disk \underline{without
interaction} is very small. It is important that if the average
parton density $ \sim | v (Y, b) |$ is large, the probability $W$
is approximately determined by the classical physics, without
significant interference effects. The average number of
interactions of the target particle with such a disk is
proportional to its density $| v (Y, b) |$.  And because these
interactions are almost independent (up to $ r_{3P} / g $), then
the distribution in a number of interactions will be close to the
Poisson form. This implies that the probability that a fast
particle does not interact at all with such a disk is
 \bel{gleik}
  W(Y,b) =  e^{-c |v(Y,b)|}       
 \ee
and  because $W (Y, b) = | S [v (Y, b) |^2$ ~this  $S(Y,b)$
directly corresponds to the Glauber eikonal~(\ref{exps}).

It is interesting to note that the minimal Glauber eikonal answer
(\ref{gleik}) is boost-invariant in the parton interaction picture
and this signals that here the t-unitarity is properly taken into
account. To see this let us consider this tunnelling process in an
arbitrary longitudinal system, when the energies of the colliding
particles are $ \sim m e^{y_1}, ~ m e^{y_2}, ~ y_1 + y_2 = Y $.
Then the probability of $ W (y_1, y_2, b) $ should depend only on
$ y_1 + y_2 $. If all individual parton interactions are not
mutually correlated, then the probability of the parton disk
$"y_1"$ to go through an another disk $"y_2" $ without interaction
on the impact parameter $b$, ~ as follows from (\ref{gleik}), will
be
 \bel{winv}
    W(y_1,y_2,b) \sim \exp{\Big( -c \int d^2 b_1 \ln
   |S(y_1,\vec{b_1})| ~\rho(\vec{b} -
                   \vec{b_1},y2) \Big) } ~\sim
 \ee
$$
\sim
  \exp{\Big(  -c \int d^2 b_1
       |~v(y_1,\vec{b}_1)~v(y_2, \vec{b} -
       \vec{b_1})| \Big)} \sim e^{-c |v(y_1 + y_2,b)|}~
       \sim W(y_1 + y_2,b)
$$
Such a non-trivial answer arises only because all quantities
entering (\ref{winv}) behave in a very special way. Namely: ~$| S
| \sim \exp{ ( -c | v | )}$~;~ the parton density in the disk $
\rho (b, y) \sim |v (y, b) |$ ~and the amplitude $v (y, b) \sim
y^{-1} \exp{(-b^2/4y + \Delta y)}$ has the diffusion structure.~
If we choose the weights $ \beta_n^2 $ in some non-Glauber form,
than the boost invariance of $W$ can be broken. It can be broken
also if we take into account the interaction between soft
pomerons, because then the saturated soft Froissart disk will be
gray.

But the expressions of the type (\ref{gleik}, \ref{winv}) take
place only if we neglect the possibility of large fluctuations in
the parton disk and consider the collision of particles only in
states with the average value of the parton density in disks given
by the amplitude $|v|$. The same will be true  whenn we take into
account only local fluctuations of the parton density around the
mean value. However, in the parton cascade very large density
fluctuations are indeed possible, even such that occupy all disk
at once, and which arise mostly from fluctuations at the first
steps of the cascade.

The largest fluctuations of this type are such that there are at
all no secondary partons in a fast particle state, so that the
state contains only the minimal valence components. Such state is
completely transparent even for arbitrary large $E$. Therefore,
here the value of $|S|^2$ is determined directly by the
probability of fluctuations bringing the fast particle to such a
transparent state. In theories with growing cross-sections of
vector type ( like QCD ) the probability to be in the state
without any secondary soft partons is $w (y) \sim \exp{ ~(-c_1
y)}$. This is just the probability that the valence partons do not
emit any soft photon (gluon ...) in the whole interval of rapidity
$y$. It is essential that such an estimate of $w (y)$ is valid for
any $y$, including the case of Froissart limit with the soft
saturated parton disk.

It is evident that the tunnelling mechanism associated with such
fluctuations is much more efficient than that corresponding to a
simplest eikonal, it is compared to a tunnelling in an ``average''
state.

The state without additional partons arise from specific
fluctuations in the initial stage of the cascade. Therefore, one
may think that the effects of these fluctuations can be taken into
account in the reggeon amplitudes, just properly adjusting the
coefficients $N_n $ in eikonal series to get
 \bel{wtun}
   | S(y, b=0) |^2 \sim w(y) \sim \exp{(-c_1 y)} ~.
 \ee
It follows from (\ref{fac}) and (\ref{pred}) that for this it is
necessary that ~$\beta_n^2 \sim n!$ ~.~ Than we will have ~$| S |
\sim 1 / | v |^a \sim \exp{(-ya \Delta)}$. ~ So we come to the
expression for $ \beta_n $ of the same limiting type that was
discussed above.

Note that for the interaction of a projectile with the composite
target consisting of  \textit{l} particles, the probability for
such a projectile to tunnel through a dense disk will be $| S |^{2
\textit{l}}$  - if we consider only a mean disk configurations and
the independent tunnelling of all its \textit{l} components. Or
this probability will contain an extra small factor $\sim r_0^{2
\textit{l}}$, where  $r_0$ is the size of the region in which all
$\textit{l}$ target particles should gather, so that their
tunnelling looked like the passage of a single particle. These
probabilities are parametrically of the same order.  At the same
time, in configurations where fast particles do not have a
secondary parton (it is no soft disk), the probability to pass
without interactions will also be the same $| S |^2 $ and does not
depend on~$\textit{l}$.

If one takes into account also hard components in the Froissart
disk (like in QCD at $E \rightarrow \infty$) then the estimate
(\ref {wtun}) will change. Now it is necessary to find the
probability that the valence components (quarks) do not emit
secondary gluons-partons with arbitrary energies and transverse
momenta. This leads (in the language of the model section 2c) to
the behavior
$$
 w(y) \sim \prod_{i=1}^{\sim \sqrt{y}} |S_i(y,b=0)|^2 ~~\sim
\exp{(-c y \ln y)}  ~,~~~~~~~
 S_i \sim e^{-y \Delta_i}~,~~    \Delta_i \simeq c/i~~.
$$
But this quantity is still much bigger than the corresponding
Glauber probability
$$
w(y) \sim \exp{(- |v|)}   \sim \exp{ \big( -c_2 y^{-1}}
   \exp{(\Delta y) \big)}~.
$$.

\subsection*{\it{ 4c~~ Remarks on the diffraction generation
     processes at very high energies } }

Here we briefly discuss how processes of diffraction generation
are connected with the structure of beams entering the $N_n$
vertices.

At very high energies, when $| v | \gg 1$, in the inelastic
vertices $N_n$ the beams with the large average number of
particles $\nu \sim n \sim | v | \gg 1$ will be significant.
Therefore, it may seem that the average number of particles
created in the diffraction generation processes can grow in the
same way - as $| v | \sim e^{\Delta y}$  -  so as grows  the
average number of cut lines in the non-planar Feynman diagrams
that determine the $N_n$ vertices.

However, as follows from the relation (\ref{difgen}), the
contribution to the cross-section of inelastic diffraction
processes comes only from such impact parameters, when $| v (y, b)
| \sim $ 1 and $ S \sim 1 $, and not from configurations when $1-
|S| \ll 1$ or $|S| \ll 1$
 \footnote {If the hard components in $v$ are also big, then the
situation may change somewhat due to the diffraction generation of
the hard jets.}.
 Therefore in the corresponding eikonal series only the several
number of terms will be non small, and this leads to a small
average multiplicity of particles in diffraction beams. Such
behavior is natural in the parton picture, because for the impact
parameters when $| v (y, b) | \gg 1 $, all valent components of
the initial state are absorbed equally and therefore give a
contribution only to the elastic diffraction component. Thus, the
diffraction generation processes come only from the border of a
dense parton disk.

The interaction between pomerons $\Pm$ slightly changes this
picture, because now there appear other types of diffraction
processes (of the $3 \Pm $... type), which give the wide
distribution over the beam masses.

In the Froissart limit the elastic  amplitude $F (Y, b) \simeq i
\tilde {\theta} (r_0^2 Y^2 - b^2) ~ $, where $ \tilde {\theta} $ -
is the theta-function with the edge, smeared on the width $y_0
\simeq 1$. In this case the dependence of the cross section from
the diffraction beam mass ($\sim e^y$) has the $5\F$ form
$$
d \sigma_{dif}(Y,y) /dy \sim Y^{-1} \sqrt{(y+y_0) (Y-y+y_0) }~.
$$
This corresponds to the growth  of the average mass of diffraction
states as  $\sim s^{1/4}$.  The full cross-sections of these
processes $ \sim Y^{3/2} $ , while the cross section of the
diffractive beams with limited mass is $ \sim Y $.

\subsection*{\it{ 4c~~ Comments on peculiarities of the nonperturbative QCD } }

We have discussed abow the properties of $N_n$ vertices that are
not QCD-specific. In the Regge approach main peculiarities of QCD
are manifested only in the specific value of parameters of Regge
theory. The perturbative QCD dynamics leads to BFKL-like structure
of pomeron and corresponds to the inclusion of the large parton
virtualities with the increasing  energy.

The main nonperturbative field-theoretic effects are firstly
reflected in the significant difference between values of regge
vertices $ g_{P}, N_2 \simeq g_P^2 $ ... describing the emission
of pomerons by the external hadrons and of vertices like $R_{3P},
... $ describing the interaction between pomerons themselves.

It is often assumed that this difference is a consequence of the
presence of two different scales in the nonperturbative QCD. The
"big" radius $\rho_1 \sim 5 GeV^{-1}$ is defined by the scale of
the quark condensate $<q\bar{q}>$, and the second ``relatively
small'' radius $\rho_2 \sim (0.5 \div 1)GeV^{-1}$ can be
associated with the value of the correlation lengths in the gluon
vacuum \cite{simonov}. The first scale determines the size of
hadrons and also the size and value of vertices $g_P, N_2,N_3,
...$. The second scale defines the QCD-boundary between
perturbative and non-perturbative regimes as well as the average
``size of gluons'' and of the corresponding (pomeron $\bigotimes$
pomeron) interaction vertices $r_{3P} = r_{P-> 2P},~r_{2P-> 2P}
~,...$
 \footnote {The smallness of $\alpha_P '$ compared to its
natural value $\sim m_{\pi}^{-2} \div m_{\rho}^{-2} $, and the
smallness of the ``experimental'' value of the pomeron intercept $
\Delta_P $ compared to the perturbative BFKL-value can also be
associated with the presence of the two QCD scales}.

The parameter $\rho_1/\rho_2$ defines the relative contribution of
the non-enhanced (eikonal-like) and enhanced (with inter-pomeron
interaction) reggeon diagrams. And the big value of this parameter
$\rho_1/\rho_2 \gg 1$ in QCD causes that the contribution of the
non-enhanced reggeon diagrams dominates in the hadron cross
sections up to very high energies, and the $3 \Pm ,...$ screening
is not very important even at LHC-energies. Only for the
interaction of heavy nuclei, when the number of the effective
pomeron exchanges increases in $A^{1/3}$ times, the fusion of the
"thin" pomeron chains (described by $r_{3P}$) becomes also
important~
 \footnote {Probably, this also explains why the quark-gluon
string model  \cite{KaTer} describes quantitatively all major
observed effects in the high-energy $hh$ interactions, and with
inclusion of simplest $3 \Pm$ tree diagram corrections also the
main $A \otimes A $ effects.}.

In such a picture the  pomeron lines, entering in $N_n$ vertices,
are attached to the valent quarks or to additional  $q \bar {q}$ ~
pairs, arising from the exchange of light quarks with the
$<q\bar{q}>$ condensate. Such  $q \bar {q}$ pars can be present in
the valent part of the Fock wave function of fast hadron with some
weights $w_k$ .

One can estimate the connection of $w_k$ with  $N_n$ identifying
these weights $w_k$ with the probabilities of the valent
constituents of Section 4a and directly using relations
(\ref{betw} - \ref{basim}). Then we come to the following type
association
$$
~~~~~~~~  N_n \sim  c^n ~~~~~~~~~~~
 \rightleftarrows~~~~~
          ~ w_m \sim c_1^m /m!
$$
$$
~~~~~   N_n \sim  c^n \sqrt{n!}
 ~~~~~~\rightleftarrows~~~~~~
                           w_m \sim c_1^m /\sqrt{m!}~.
$$
Considering the $w_k$ as a distribution of the number of $ q
\bar{q}$ ~ pairs in the nucleons wave function, one can estimate
its behavior in various models, for example in the model of
``bag'', whose size is fluctuating due to exchange of quarks with
the  surrounding $<q\bar{q}>$ condensate. Probably, this mechanism
can be effectively taken into account through the contributions to
$N_n$ from the tree diagrams with $ \pi $ and $ \rho $ reggeons
(see Section 5).

\subsection*{\it{ 4d~~
Comments about the specifics of high energy gravitational
interactions}}

The primary amplitude  for a gravitational exchange increases very
fast with energy $v \sim s / m_p^2 $~, where $m_p$ is the Planck
mass. But for the ``existing'' energies such amplitudes are too
small $| v | \ll 1$. The higher eikonal contributions (multiple
gravitational exchanges) can become important only at planckian
energies, where $ s > m_p^2 $ ~ or for collisions of an
ultra-relativistic particle with the energy $\varepsilon$ with a
heavy object of mass $M$, when the effective $s \sim 2 \varepsilon
M > m_p^2 $.

For transplankian particles when $ \varepsilon \gg m_p $, one can
expect that all partons in the Fock wave function  will have
limited energies $ \omega_i \lesssim m_p$ . ~Because of this, in
the beam decomposition (\ref{decom}), (\ref{Noperat}) of the
vertices $N_n$ the many-particle beams with the average number of
particles $n \sim \varepsilon / m_p$ ~ could dominate. For this
reason, one can expect that the value of $N_n$ will increase with
the $n \gg 1$ much faster than for the ordinary eikonal.

But this does not take place due to the compensation between the
contribution to $N_n$ from the growth with $ \varepsilon$ of the
number of particles in beams and by the reduction of effective
interaction constants of the individual particles in the beam.

This can be easily seen by considering the general properties of
the beam vertices matrix elements of $\langle \nu_1 | \GG
(k_{\bot}) | \nu_2 \rangle$ ~, which enter in (\ref{Noperat}). In
the gravitational case they are proportional to the matrix
elements of $T_{- -}$ components  of energy-momentum  $T_{\mu
\nu}$. And these values are determined only by the energy of
states $| \nu_i \rangle$ ~, equal to the full energy $ \varepsilon
$ of a fast particle. The relevant value of the $\varepsilon_1 *
\varepsilon_2$ coming from the product of two connected $\GG$
vertices are included to the factor $s/m_p$ , attributable to each
graviton exchange. Therefore, in the case of gravity , the
additional factors in $N_n$ due to an increase with $n$ of the
average number of particles in beams do not appear, and the
unitarized amplitude remains fairly close to the minimal Glauber
form.

This could be explained a little bit differently.  Because of the
universality of gravitational interaction, it does not change the
relative weights of various components of the Fock wave function
at the zero momentum transfer. Therefore, at $k_{\bot i} = 0$  all
the off-diagonal contributions to vertices  $ \langle \nu_i | \GG
(k_{\bot i}) | \nu_{i + 1} \rangle $ disappear and
 \bel{gravn}
  N_n (k_{\bot i} \sim 0 ) \simeq   m_p^{-n}
   \prod_{i=1}^n (1 + (\rho_n k_{\bot i})^2 + ...)~,~
 \ee
where the quantities $\rho_n \sim \rho_0 =$  the average
transverse size of a fast particle  $\rho_0 \sim 1 / m$.

It is possible that a certain growth of $N_n$ in (\ref{gravn}) can
take place when $k_{\bot i} \ne 0 $, due to a statistical
multiplication of beams, similar to the hadron case, so that
$\rho_n \sim  \rho_0 \sqrt{n}$, and it may again result in a
growth of $ N_n (k_{\bot i} \sim m) \sim  \sqrt{n!}$.

It is interesting to note that the structure of eikonal vertices
for gravitons, given by (\ref{gravn}), resemble the case of a week
coupling \cite{GribovVK} in reggeon theory.

\vspace{5mm}
\section*{\bf 5.~~
Behavior of $ N_n $ vertices for large $ n $}

\vspace{2mm}

If the "primary"  amplitude  $ v (y, b) $ increases with energy,
then  the higher order  terms in the eikonal  series (\ref {sexp})
become more and more significant. For the simplest Glauber eikonal
the mean $\langle n \rangle \sim | v | $. ~A more accurate
characteristics of the essential  $\langle n \rangle$ is given by
the average number of soft jets (cut pomerons) in the full
inelastic cross-section (\ref {njet}) as a function of full
rapidity $y$ and the impact parameter $b$ given by $\langle \nu
(y, b) \rangle =  2$ Im $ v (y, b)$ . The value of this $\langle
\nu \rangle$ does not depend on the value of parameters  $ \beta_n
$. ~ At the same time, the structure of distribution of $\nu$
(with the average $\langle \nu (y, b) \rangle$) depends on the
behavior of quantities $\beta_n$, and the tail of the distribution
over $\nu$ is directly related to the behavior of $\beta_n$ for $n
\gg \langle \nu \rangle$. This, in turn, determines the behavior
of the distribution of multiplicity of hadrons at $n_h \gg
\bar{n}_h$.

As follows from (\ref{bet}) the behavior of $\beta_n $ for $ n \gg
1 $ depends on  properties of the spectrum $\varphi (\tau)$ at
large $\tau $. Hence, for example, for the power asymptotic of
spectrum of type $ \ln {\varphi (\tau_0)}) \sim - \lambda \tau ^a
$ ~, one can estimate the asymptotic behavior of the eikonal
coefficients as $\beta_n \sim (n!)^{1/a}$.~ If the beam spectrum
is cut off at large $\tau$ so that $\varphi (\tau) = 0$ for $
\tau> \tau_1 > 1 $, we obtain the asymptotic behavior of the
quasi-eikonal type $ \beta_n \sim \tau_1 ^ {n +1} / n ~ $. For the
finite matrix models, we also  have eikonal-like asymptotic $
\beta_n \sim \tau_{max}^n $, where $ \tau_{max} $ is the
upper value of the spectrum. \\

To  estimate the asymptotic of $N_n$ at large $n$ it is convenient
to write expressions for $N_n$ in the symbolic form, as in
(\ref{Noperat}), where the averaging corresponds to the summation
over  states of particles in beams. Next, for a qualitative
estimate one can accept that the main contribution to the matrix
elements of vertices $\GG_i $ comes from almost diagonal
transitions, and also perform averaging separately for each vertex
$\GG$. Then
$$
 N_n ~=~ \langle  P_{in} | \GG (k_1)
     \GG (k_2) ... \GG (k_n) | P_{out} \rangle~
  ~\simeq~   \prod_{i=1}^{n}  \langle  \GG_i \rangle \simeq
         \prod_{i=1}^{n/2} \langle  \GG (\nu_i)\rangle^2 ~,
$$
where $\nu_i$ is the average number of particles in  i-beam. If we
also take that
$$
 \langle \GG (\nu)  \rangle \sim  \nu \cdot g ~,
$$
that is we assume that each pomeron interacts  independently with
all $\nu_i$ particles in beam,  then we get
  \bel{npuch}
    N_n \equiv g^n \beta_n \sim g^n
      \prod_{i=1}^{n/2} \nu^2_i  \sim  g^n
    \exp{ \big(~ 2\int_0^{n/2}  d i~\ln \nu_i  ~\big) }~.
  \ee
It can be expected that for large $n$  the growth of average
numbers of particles in the sequence of these beams is
approximately statistical. So, when we move in $N_n$ along the
sequence of such beams - from edge (incoming particles) to the
``middle'' beam, the values of $\nu_i$ vary on average as $\nu_i
\sim i^a, ~ ~ a \simeq 1/2 $. Then we get from (\ref {npuch}),
that the vertices $N_n$ depend on $n$ as
 \bel{asym1}
 \beta_n \sim  e^{a n (\ln n  -1)} \sim (n!)^a~.
 \ee
Such an increase for large $n$ is much faster then the Glauber
case $ \beta_n \sim c^n $, and for the simple diffusion case when
$a = 1/2$, gives the same answer $\beta_n^2 \sim n!$ as
(\ref{fac}).\\

It is also possible to estimate the behavior of $N_n $ for large
$n$ in another way - considering the t-channel contributions to $
N_n $, coming from various reggeon diagrams with non-vacuum
reggeons. Such a description can be regarded as a t-dual
representation with respect to the s-channel beam pictures
 \footnote {Probably the behavior of $N_n$ vertices of such a type can
 be expected in string and dual models}.
\begin{figure}[h]
\begin{center}
\includegraphics[scale=0.55, keepaspectratio=true]{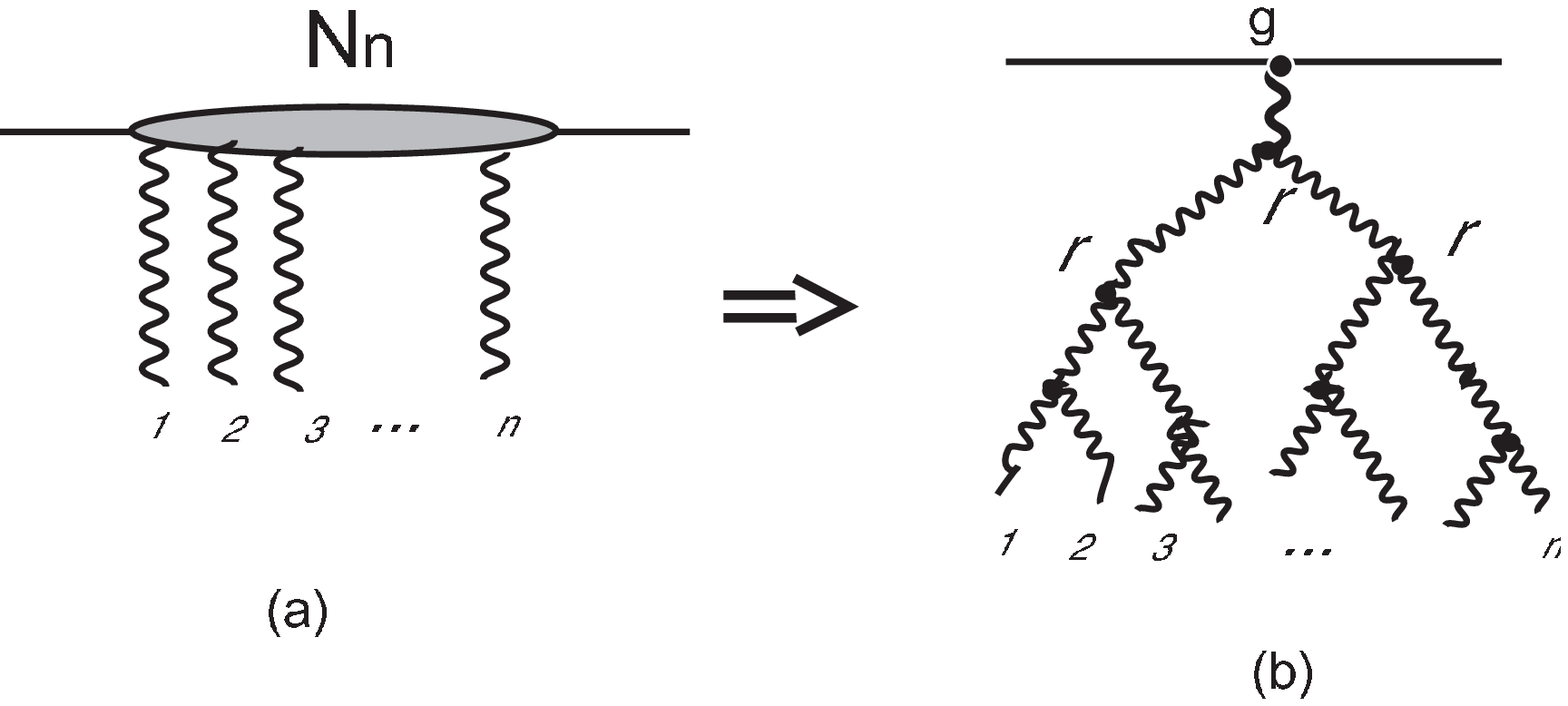}
\parbox{12cm}{    {\bf Fig.3~:}~~
 {\bf (a)} ~Diagram for $N_n$ vertex.
 ~~~{\bf (b)}  Tree diagram for $N_n$ with nonvacuum reggeons .
 } \label{Fig3}
\end{center}
\end{figure}
For simplicity let us consider only the tree diagrams with
reggeons $R \neq \Pm$. The contribution to $N_n$ from such
diagrams  can be obtained from the recursion relation
 \bel{recr}
    N_{n+1}  \simeq  r n N_n~,~~~~N_1 = g ~,
 \ee
where $r$ is the value of $3 R$ vertex, and all rggeon lines on
the diagrams are taken at a certain average value of the
transverse momentum. The solution of (\ref{recr}) has the form
$$
      N_n \sim  g r^{n-2}  n!
$$
However, such a rapid growth of $N_n$ is probably not possible for
very large $n$, since in the corresponding sequence of s-channel
beams, when reaching a certain critical particle density, the
saturation occurs. This, being applied to the recursion relations
(\ref{recr}), may be manifested in a condition that after some $n
> n_{crit}$ one can attach additional reggeons only to the border
(in the transverse plane) of the beam-disk, where the density is
not yet saturated, and that contains only $\sim \sqrt{n}$ partons
 \footnote{The same phenomenon can take place also if in the sum of
t-channel diagrams for $N_n$ we include as well all diagrams with
loops, that effectively take into account the R-reggeon gluing,
leading to the R-saturation.}.
 Therefore, the relation (\ref {recr})  is effectively
replaced by
$$
  N_{n+1}  \sim  r \sqrt{n} N_n~~,
$$
and now it has a solution
 \bel{asym2}
  N_n \sim \sqrt {n!}
 \ee
So we come again to the fast asymptotic behavior of  vertices
$N_n$ of the same type as (\ref{asym1}, \ref{fac}).\\

Some information about the behavior of vertices $N_n $ at large
$n$ can be obtained considering the energy behavior of the
probability that a particle with the high energy $\sim \exp{( y
)}$ be in a Fock state with the minimal possible number of
partons. This amplitude defines the behavior of the cross-section
for a hard elastic scattering of hadrons \cite{larget}, and it is
expressed by the elastic $S(y, b = 0)$ matrix at the zero impact
parameter. It can be seen ( Section 4b) from the comparison
between the regge and the parton pictures (this is in fact a
constraint from the s-unitarity) that we should have $S(y, b = 0)
\sim \exp{ (- c y)}$, and this corresponds to the growth of $N_n^2
\sim n!$.

\vspace{5mm}

\section*{\bf 6. Conclusion}

One of the goals of this work is to draw attention to the fact
that the most simple and very popular Glauber eikonal
unitarization can incorrectly adjust the amplitude of process
especially in the case when the primary cross-section increases
with energy. Although this minimal method leads to amplitudes that
already satisfy some conditions following  from  ~s-unitarity, it
is probably not consistent with other conditions and with
t-unitarity.\\

It can be expected that the weights of multiple exchanges grow
significantly faster than in the case of the Glauber eikonal, and
it is very likely that they grow extremely fast $ (\sim n! ~) $  ~
In these conditions the unitarized S-matrix can be asymptotically
closer to the form $ ~ S \sim i / v $ ~, and not to $ ~ S \sim
\exp{ (iv) }$, as in the case of Glauber eikonal.\\

The adequate choice of the s-channel unitarization of elastic
amplitudes is also important in a ``practical sense'' - in
connection with various phenomenological models and generators.
The weights of the multiple exchanges enter almost all hadron
amplitudes at high energies, and so they affect directly the
behavior of cross-sections of hadron processes. Examples include
the pattern of growth with energy of inclusive cross sections and
the shape of the multiplicity distribution of produced particles.

\vspace{10mm}
\nin {\bf ACKNOWLEDGMENTS} \\
\nin I thank K.G.~Boreskov and I.M.Dremin for conversations and
useful comments.~~

\nin The financial support of CRDF through the grant
RUP2-2961-MO-09 is gratefully   acknowledged.

\vspace{4mm}


\end{document}